**SHORT COMMUNICATION**

# Overcoming challenges of translating deep-learning models for glioblastoma: the ZGBM consortium

[1]HARIS SHUAIB, [2]GARETH J BARKER, PhD, [2]PETER SASIENI, PhD, [2]ENRICO DE VITA, PhD, [2]ALYSHA CHELLIAH, [3]ROMAN ANDREI, [4]KEYOUMARS ASHKAN, [5]ERICA BEAUMONT, [1]LUCY BRAZIL, [6]CHRIS ROWLAND-HILL, [4]YUE HUI LAU, [2]AYSHA LUIS, [7]JAMES POWELL, [1]ANGELA SWAMPILLAI, [8]SEAN TENANT, [9]STEFANIE C THUST, [9]STEPHEN WASTLING, PhD, [1]TOM YOUNG, [2,4]THOMAS C BOOTH, MA MRCP FRCR PhD and ZGBM consortium

[1]Guy's & St Thomas' NHS Foundation Trust, King's College, London, United Kingdom
[2]King's College London, London, United Kingdom
[3]The Oncology Institute "Prof. Dr. Ion Chiricuță" Cluj-Napoca, Cluj-Napoca, Romania
[4]King's College Hospital NHS Foundation Trust, London, United Kingdom
[5]Lancashire Teaching Hospitals NHS Foundation Trust, Lancashire, United Kingdom
[6]Hull & East Yorkshire Hospitals NHS Trust, England, United Kingdom
[7]Velindre University NHS Trust, Wales, United Kingdom
[8]The Christie NHS Foundation Trust, Manchester, United Kingdom
[9]National Hospital for Neurology and Neurosurgery, UCL Institute of Neurology, London, United Kingdom

Address correspondence to: Dr Thomas C Booth
E-mail: *tombooth@doctors.org.uk*

**Objective:** To report imaging protocol and scheduling variance in routine care of glioblastoma patients in order to demonstrate challenges of integrating deep-learning models in glioblastoma care pathways. Additionally, to understand the most common imaging studies and image contrasts to inform the development of potentially robust deep-learning models.
**Methods:** MR imaging data were analysed from a random sample of five patients from the prospective cohort across five participating sites of the ZGBM consortium. Reported clinical and treatment data alongside DICOM header information were analysed to understand treatment pathway imaging schedules.
**Results:** All sites perform all structural imaging at every stage in the pathway except for the presurgical study, where in some sites only contrast-enhanced $T_1$-weighted imaging is performed. Diffusion MRI is the most common non-structural imaging type, performed at every site.
**Conclusion:** The imaging protocol and scheduling varies across the UK, making it challenging to develop machine-learning models that could perform robustly at other centres. Structural imaging is performed most consistently across all centres.
**Advances in knowledge:** Successful translation of deep-learning models will likely be based on structural post-treatment imaging unless there is significant effort made to standardise non-structural or peri-operative imaging protocols and schedules.

There have been promising advances in the medical applications of artificial intelligence (AI) computer vision methods, such as using deep learning, which have the potential to improve the clinical management of glioblastoma patients.[1] A challenge for the clinical translation of existing research has been the paucity of large-scale external validation of these methods, as well as the reliance on advanced imaging techniques that are not commonly acquired in routine practice, such as perfusion MRI.[2] This has meant that to date there has been no clinical translation of these applied imaging methods for glioblastoma patients, despite similar techniques becoming increasingly common in neurological conditions such as stroke and employed in randomised-controlled trials.[3,4]

The ZGBM (zeugmatography for glioblastoma) consortium is a collaboration of leading neuro-oncology centres across the UK that is working to address these challenges, and thus improve the treatment of glioblastoma patients. The consortium currently consists of 16 NHS Trusts across England, Scotland and Wales and has collected a retrospective dataset of over 500 patients. It is prospectively recruiting patients newly diagnosed with glioblastoma and undergoing the Stupp treatment regimen.[5] The target recruitment for the prospective cohort is 350 patients, however, based on the consortium's current recruitment rate, total recruitment is expected to exceed the target by recruitment end on 22/05/2022.

The retrospective cohort will be used to develop AI models that are able to differentiate progression from pseudoprogression as well as understand the evolution of tumour undergoing imaging follow up after treatment. These AI models will then be tested in the prospective external cohort in order to measure their performance in a totally independent dataset, and thereby allowing us to ensure that accuracy remains high in glioblastoma patients who have undergone imaging using different MRI scanner manufacturers and scanning parameters.

The research involves three key innovations that will aid the translation of these methods into clinical practice:

- Restricting the input data to structural imaging data that is routinely acquired across the NHS and integration of image normalisation techniques to further reduce the impact of imaging variation between institutions.
- Integration of treatment data including the radiotherapy treatment plan in evaluating imaging phenotypes, as prior work has shown that there is a correlation between radiotherapy dose and MRI signal.[6]
- Development of the model as a MONAI Deploy application[7] that could, once validated, be easily integrated into the clinical workflow.

The following institutions are members of the ZGBM consortium (Figure 1) and have committed to contributing longitudinal MR images as well as treatment and clinical information: King's College Hospital NHS Foundation Trust, Guy's & St Thomas' NHS Foundation Trust, NHS Tayside, The Christie NHS Foundation Trust, Hull University Teaching Hospitals NHS Trust, Lancashire Teaching Hospitals NHS Trust, University Hospitals Sussex NHS Foundation Trust, Newcastle Hospitals NHS Foundation Trust, The Royal Marsden NHS Foundation Trust, Velindre University NHS Trust, Leeds Teaching Hospitals NHS Trust, University College London Hospital NHS Foundation Trust, Imperial College London Hospital NHS Foundation Trust, Barts Health NHS Trust, Nottingham University Hospitals NHS Trust and University Hospitals Plymouth NHS Trust.

The consortium has collected and pooled an initial set of prospective imaging data to understand the degree of variation in imaging protocols (including scheduling) across the member institutions. For illustrative purposes, a sample of five subjects per site was investigated to understand the differences in imaging protocol between sites, and patients. Five sites were included in this initial study. Table 1 details the imaging studies performed on glioblastoma patients at each of these sites, as part of their routine care, categorised in terms of their relation to treatment milestones *e.g.*, surgery or end of chemo-radiation, as well as the image contrasts acquired.

We can understand from the data presented in Table 1, that there is large variation in the schedule of imaging, as expected.[8] For example, site two only acquire imaging at the presurgical and surveillance stages in contrast to site four who acquire imaging at presurgical, postsurgical and pre-treatment stages as well as the surveillance stage.

There is also large variation in scan protocol across both sites and study time points, as expected.[8] The majority of sites perform all structural imaging (T1, T1POST, T2, FLAIR) at every time point except for the presurgical study, where in some sites only T1POST is performed. Beyond structural imaging, there is very little non-structural imaging performed and where it is performed it is not performed regularly, again as expected.[8] Diffusion MRI is the most common non-structural imaging type, being performed at every site. One site performs susceptibility-weighted imaging (SWI) more regularly than any other site, whilst two sites performed none at all. Advanced MRI techniques, including perfusion imaging (dynamic susceptibility contrast (DSC), dynamic contrast enhancement (DCE), arterial spin labelling (ASL), is even less common with only three out of six sites having performed any perfusion imaging.

From these early results using MR images from real-world ('pragmatic') imaging protocols and follow-up schedules across the UK, it is clear that one of the core challenges in developing deep learning models for glioblastoma will be the lack of standardised imaging. In order to overcome these challenges, it is key for future efforts to focus on commonly acquired images (such as structural images) that are available across the UK and are not dependent on pre-treatment imaging (including imaging during the perioperative period) as such scans are scarce in comparison to follow-up surveillance imaging.

In summary, the ZGBM consortium represents a diverse group of institutions delivering care to glioblastoma patients, which aims to prospectively validate deep learning techniques to inform and improve the management of these patients. While we have already begun to evaluate some of the MRI scheduling and imaging protocol variations across the UK, we are keen to grow our consortium to ensure that our results are translatable across the NHS as soon as possible, and as such welcome additional members and contributors.

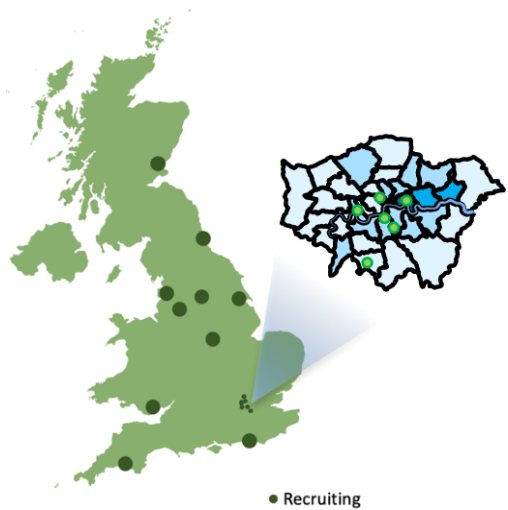

Figure 1. Map of ZGBM consortium sites.

Table 1. Counts of total studies as well as total image contrasts per type for five sites (1-5)

| | Total exams | T1POST (n) | T2 (n) | T1 (n) | FLAIR (n) | SWI (n) | DWI (n) | DSC (n) | DCE (n) | ASL (n) | T1MAP (n) | MRS (n) |
|---|---|---|---|---|---|---|---|---|---|---|---|---|
| **1** | **20** | **20** | **16** | **16** | **17** | | **16** | **1** | | | | |
| Diagnostic | 1 | 1 | 1 | 1 | 1 | | 1 | 1 | | | | |
| Presurgical | 5 | 5 | 2 | 2 | 2 | | 2 | | | | | |
| Postsurgical | 2 | 2 | 2 | 2 | 2 | | 2 | | | | | |
| Pretreatment | 2 | 2 | 2 | 2 | 2 | | 2 | | | | | |
| Surveillance | 5 | 5 | 5 | 5 | 5 | | 5 | | | | | |
| Unknown | 5 | 5 | 4 | 4 | 5 | | 4 | | | | | |
| **2** | **19** | **19** | **16** | **16** | **16** | | **16** | **1** | | | | |
| Presurgical | 4 | 4 | 1 | 1 | 1 | | 1 | 1 | | | | |
| Surveillance | 15 | 15 | 15 | 15 | 15 | | 15 | | | | | |
| **3** | **17** | **17** | **17** | **16** | **17** | **13** | **17** | **2** | | **1** | | |
| Presurgical | 5 | 5 | 5 | 5 | 5 | 4 | 5 | | | | | |
| Postsurgical | 6 | 6 | 6 | 5 | 6 | 6 | 6 | | | | | |
| Surveillance | 1 | 1 | 1 | 1 | 1 | 1 | 1 | 1 | | 1 | | |
| Unknown | 5 | 5 | 5 | 5 | 5 | 2 | 5 | 1 | | | | |
| **4** | **12** | **12** | **11** | **11** | **11** | | **11** | **1** | | | | |
| Presurgical | 3 | 3 | 2 | 2 | 2 | | 2 | | | | | |
| Postsurgical | 1 | 1 | 1 | 1 | 1 | | 1 | | | | | |
| Pretreatment | 2 | 2 | 2 | 2 | 2 | | 2 | | | | | |
| Surveillance | 6 | 6 | 6 | 6 | 6 | | 6 | 1 | | | | |
| **5** | **24** | **24** | **16** | **16** | **15** | **5** | **16** | **5** | **5** | | **4** | **1** |
| Diagnostic | 2 | 2 | 2 | 2 | 2 | | 2 | | | | | |
| Presurgical | 8 | 8 | | | | | | | | | | |
| Postsurgical | 1 | 1 | 1 | 1 | 1 | | 1 | | | | | |
| Surveillance | 13 | 13 | 13 | 13 | 12 | 5 | 13 | 5 | 5 | | 4 | 1 |
| **Grand Total** | **92** | **92** | **76** | **75** | **76** | **18** | **76** | **10** | **5** | **1** | **4** | **1** |

ASL, arterial spin labelling; DCE, dynamic contrast enhanced; DSC, dynamic susceptibility contrast; DWI, diffusion-weighted; FLAIR, fluid-attenuated inversion-recovery; MRS, MR 1H-Spectroscopy; SWI, susceptibility weighted; T1, T1-weighted; T2, T2-weighted; T1MAP, T1 map; T1POST, gadolinium-enhanced.

**ACKNOWLEDGEMENTS**

The following contributors are part of the ZGBM consortium. The authors would like to acknowledge their contribution to the paper: Juliet Brock, MD, Brighton and Sussex University Hospitals NHS Trust; Stuart Currie, PhD, Leeds General Infirmary; Kavi Fatani, Leeds General Infirmary; Karen Foweraker, Nottingham University Hospitals NHS Trust; Jennifer Glendenning, Maidstone and Tunbridge Wells NHS Trust; Nigel Hoggard, Sheffield Teaching Hospitals NHS Foundation Trust; Avinash K Kanodia, MD, Ninewells Hospital; Anant Krishnan, PhD, Barts Health NHS Trust; Mark DV Thurston, University Hospitals Plymouth NHS Trust, University of Plymouth; Joanne Lewis, Northern Centre for Cancer Treatment, Newcastle upon Tyne; Christian Linares, Guy's & St Thomas' NHS Foundation Trust; Ryan K Mathew, Leeds Teaching Hospitals NHS Trust, University of Leeds; Satheesh Ramalingam, University Hospitals Birmingham NHS Foundation Trust; Vijay Sawlani, University Hospitals Birmingham NHS Foundation Trust; Liam Welsh, The Royal Marsden NHS Foundation Trust; Matt Williams, PhD, Imperial College Healthcare NHS Trust, Computational Oncology Imperial College.

**FUNDING**

This work is funded by an NIHR Doctoral Research Fellowship. This work is also supported by a Radiology Research Trust grant, King's College Hospital Research and Innovation, and the Wellcome/Engineering and Physical Sciences Research Council Centre for Medical Engineering (WT 203148/Z/16/Z).